\documentclass[%
superscriptaddress,
notitlepage,
twocolumn,
 amsmath,amssymb,
 aps,
prl,
]{revtex4-1}

\usepackage{graphicx}
\usepackage{dcolumn}
\usepackage{bm}
\usepackage{hyperref}

\usepackage{xcolor}
\usepackage[normalem]{ulem}

  \newcommand{\av}[1]{\left\langle#1\right\rangle}
  \newcommand{\cbr}[1]{\left(#1\right)}
  \newcommand{\sbr}[1]{\left[#1\right]}
  \newcommand{\bbr}[1]{\left\{#1\right\}}

\DeclareRobustCommand{\vect}[1]{
  \ifcat#1\relax
    \boldsymbol{#1}
  \else
    \mathbf{#1}
  \fi}

\begin{document}

\preprint{}

\title{Control of Active Brownian Particles: An exact solution}

\author{Marco Baldovin}
\affiliation{Universit\'{e}  Paris-Saclay,  CNRS,  LPTMS,  91405,  Orsay,  France}%
 \affiliation{Institute for Complex Systems, CNR, 00185, Rome, Italy}
 \email{marco.baldovin@cnr.it}
\author{David Gu\'{e}ry-Odelin}%
\affiliation{Laboratoire  Collisions,  Agr\'{e}egats,  R\'{e}eactivit\'{e}e,  FeRMI,  Universit\'{e}  de  Toulouse,  CNRS,  UPS,  France}%
\author{Emmanuel Trizac}%
\affiliation{Universit\'{e}  Paris-Saclay,  CNRS,  LPTMS,  91405,  Orsay,  France}
\affiliation{Ecole normale sup\'erieure de Lyon, F-69342 Lyon, France}%

\date{\today}

\begin{abstract}

Control of stochastic systems is a challenging open problem in statistical physics, with a wealth of potential applications from biology to granulates. Unlike most cases investigated so far, we aim here at controlling a genuinely out-of-equilibrium system, the two dimensional Active Brownian Particles model in a harmonic potential, a paradigm for the study of self-propelled bacteria.  We search for
protocols for the driving parameters (stiffness of the potential and activity of the particles) bringing the system from an initial passive-like steady state to a final active-like one, within a chosen time interval. The exact analytical results found for this prototypical model of self-propelled particles brings control techniques to a wider class of out-of-equilibrium systems.

\end{abstract}

\maketitle

\textit{Introduction --- } Active matter is one of the most studied and promising topics of out-of-equilibrium statistical physics~\cite{elgeti2015physics,bechinger2016active, gompper20202020,OByrne2022}.  Inspired by the behaviour of biological systems such as bacteria and cells, this class of problems is characterized by the presence of internal mechanisms (e.g., self-propulsion) inducing nonzero entropy production, through energy dissipation. Motility-induced phase separation~\cite{cates2015motility,fily2012athermal}, pattern formation~\cite{digregorio2018full,farrell2012pattern} and velocity self-alignment~\cite{caprini2020spontaneous} are typical hallmarks of the intrinsic out-of-equilibrium nature of these systems. Among the others, activity is a key future of nano-swimmers~\cite{Golestanian_2007}, complex colloidal or bacteria dynamics~\cite{Howse2007,Gejji2012}, and active transport~\cite{Bressloff_2013}.
 While the engineering of such systems becomes possible \cite{OByrne2022},
it remains a challenge to control activity in general. This demands a proper understanding of the dynamics under confinement, an important endeavour for active objects~\cite{pototsky2012active,DauchotDemery2019}. The present work is a step in this direction.

Several experiments have shown the possibility to tune the degree of activity of active matter~\cite{buttinoni2012active,maggi2015micromotors,vizsnyiczai2017light,vutukuri2020light,Militaru2021}. In Ref.~\cite{buttinoni2012active}, for instance, silica spheres of a few $\mu m$ radius, partly covered by chromium and gold (Janus particles) are diluted in a binary mixture of water and 2,6-lutidine, that reacts with the surface of the particles and induces self-propulsion. The reaction is tuned by the intensity of light, so that the persistent velocity can be controlled.
This light-dependent tuning is a promising mechanism for the control of active fluids and may have useful applications, e.g., for the clogging/unclogging of microchannels~\cite{Dressaire2016CloggingOM,caprini2020activity}. The main idea behind these applications is to bring the system from a passive-like to an active-like phase, and vice-versa, and to take advantage of the different distribution of the particles in the two states.

The time needed to switch the system from one phase to the other will depend, in general, on the protocol that is employed to change the values of the controlling parameters. A sudden change of the external light, for instance, may then require a long time for the 
relaxation
of the system to the desired final distribution. It is thus important to search for protocols that allow to execute the transition in a controlled way, in a short time. This type of problems, that can be subsumed under the terminology of ``swift state-to state transformations'' (SST)~\cite{guery2022driving}, has witnessed a surge of interest in the last 15 years. The first studies are in the realm of quantum mechanics~\cite{torrontegui2013shortcuts}, where they are referred to as ``shortcuts to adiabaticity''; applications to statistical physics and stochastic thermodynamics are more recent~\cite{martinez2016engineered,guery2022driving}.

In this letter, we study such SST problems for a system of Active Brownian Particles (ABP) in two dimensions~\cite{romanczuk2012active, pototsky2012active, solon2015active, solon2015pressure,basu2018,Basu2019, caporusso2020motility}. This is one of the simplest and most used models mimicking the behaviour of self-propelled particles like bacteria~\cite{romanczuk2012active}, whose fluctuating hydrodynamics has been shown to be equivalent to the Run-and-Tumble model describing the above mentioned Janus particles~\cite{cates2013active, solon2015active}. We will assume that the system is confined in an external harmonic potential with tunable stiffness, as done for instance in Ref.~\cite{takatori2016acoustic} by using acoustic waves. The stationary (steady-state) distribution of this model was found in Ref.~\cite{Malakar2020Steady}.
With these assumptions, we will describe a class of analytical protocols leading the system from a passive-like to an active-like state with the same stiffness in a finite time, and vice-versa. Among this class of control protocols, we will identify the one minimizing the total time required for the transition.

\textit{Model --- } 
The state of 2 dimensional ABP is defined by a spatial position $\vect{\rho}=(\rho \cos \varphi,\rho \sin \varphi)$ for the center of mass, and an angle $\theta$ associated to the orientation of the particle. The particle's velocity is given by the sum of a self-propulsion contribution along the direction of $\theta$,  $\vect{e}(\theta)$, with constant modulus $u_0$, plus a thermal Gaussian noise. The orientation $\theta$ is also subject to Gaussian fluctuations. In addition, the effect of an external potential will be taken into account. We consider the case of isotropic harmonic confinement, resulting in a force $-k \vect{\rho}$ pointing toward the origin ($k$ being the stiffness). In the overdamped limit, the time evolution is then given by the coupled Langevin equations
\begin{equation}
\begin{cases}
\frac{d \vect{\rho}}{d \tau}=&u_0\widehat{\vect{e}}(\theta)-\mu k \vect{\rho}+\sqrt{2 D_t} \, \vect{\xi}_r(\tau),\\ 
\frac{d \theta}{d \tau}=&\sqrt{2 D_{\theta}} \, \vect{\xi}_{\theta}(\tau)\,,
\end{cases}
\end{equation}
where $\tau$ is the time, $\mu$ stands for the mobility, $\vect{\xi}_r(\tau)$ and $\vect{\xi}_{\theta}(\tau)$ are Gaussian white noises, while $D_t$ and $D_{\theta}$ are the translational and the rotational diffusivities.

 In the following we will consider the dimensionless variables $t= D_{\theta} \tau$, $\vect{r}=\vect{\rho}/\Delta$, where $\Delta$ is a unit of length to be specified, and the dimensionless parameters
\begin{equation}
\label{eq:dimlessparam}
    \kappa=\frac{\mu k}{D_{\theta}}\,,\quad \lambda=\frac{u_0}{D_{\theta}\Delta}\,,\quad \alpha=\frac{D_{\theta}\Delta^2}{D_t}\,.
\end{equation}
The stationary probability density function (pdf) for this problem can be worked out analytically as a series expansion in powers of $\lambda$~\cite{Malakar2020Steady}:
\begin{equation}
\label{eq:statpdf}
\begin{aligned}
 P_{s}(r,\chi)&=F(r,\chi|\kappa, \lambda,\alpha)\\&\equiv\alpha\sum_{m=0}^{\infty}\lambda^m \alpha^{m/2}\sum_{2n+|l|=m}C_{n,l}^{(m)}(\kappa)\phi_{n,l}(\sqrt{\alpha}r,\chi|\kappa)\,,
 \end{aligned}
\end{equation}
where $r=|\vect{r}|$ and $\chi=\theta-\varphi$, $\varphi$ being the orientation of the vector $\vect{r}$. Here, the $C_{n,l}^{(m)}$ are coefficients that can be determined by suitable recursive rules, while the explicit expression of $\phi_{n,l}$ is known~\cite{Malakar2020Steady}; their definition is recalled in the Supplemental Material (SM)~\cite{SM}. In the following we choose $\Delta=\sqrt{D_t/D_{\theta}}$ as the rescaling length, so that $\alpha=1$. The stationary pdf then depends only on $\kappa$ and $\lambda$, that have the respective meaning of a dimensionless stiffness and of a normalized persistence length accounting for the degree of activity of the system.

\textit{Swift state-to-state transformations --- } We now face the problem of bringing the system from an initial stationary state, characterized by $\lambda=\lambda_i$, to a final state $\lambda=\lambda_f$ with the same $\kappa=\kappa_0$, in a given time interval $t_f$. Solutions obtained as sequences of stationary states (the so-called ``quasi-static'' protocols), where the control parameter $\lambda(t)$ is slowly varied between $\lambda_i$ and $\lambda_f$, require infinite time. They are not suitable for our purposes, since we wish to complete the connection in a given finite time $t_f$. Instead, we will take advantage of the known steady-state distribution~\eqref{eq:statpdf} to look for time-dependent, non-quasi-static protocols. In particular, we search for an exact solution with functional form:
\begin{equation}
\label{eq:pdfansatz}
P(r,\chi,t)\equiv F(r,\chi|\widetilde{\kappa}(t),\widetilde{\lambda}(t),\widetilde{\alpha}(t))\,,
\end{equation} 
where the functions $\widetilde\kappa$, $\widetilde\lambda$ and $\widetilde\alpha$, yet to be specified, completely define the instantaneous state of the system (i.e., the pdf of the active particle).  We introduce the tilde
variables in order to make a clear distinction between the control
parameters ($\kappa$ and $\lambda$) and the variables describing the state of the
system ($\widetilde{\kappa}$, $\widetilde{\lambda}$ and $\widetilde{\alpha}$). The former are directly
controlled during the experiments: in the proposed experimental setup, see next paragraph,
they would be the (rescaled) stiffness of the external potential and
the (rescaled) persistence length induced by the chosen light
intensity. The tilde variables, instead, define the probability
density function of the active particle at a given instant of time,
which evolves in turn according to the Fokker-Planck equation defined
by $\kappa$ and $\lambda$. While tilde and non-tilde variables coincide in the
stationary state, they are different during a dynamic evolution.

 We require $\widetilde{\kappa}(t)$, $\widetilde{\lambda}(t)$ and $\widetilde{\alpha}(t)$ to be continuous, positive functions of time such that 
\begin{subequations}
\label{eq:cond}
\begin{eqnarray}
    \widetilde{\kappa}(0)&=\kappa_0\quad \quad\widetilde{\kappa}(t_f)&=\kappa_0\,,\\
    \widetilde{\lambda}(0)&=\lambda_i\quad \quad\widetilde{\lambda}(t_f)&=\lambda_f\,,\\
\label{eq:condalpha}
 \widetilde{\alpha}(0)&=1\quad\quad\widetilde{\alpha}(t_f)&=1\,,
\end{eqnarray}
\end{subequations}
so that at the beginning and at the end of the process the system is in a stationary state induced by the external parameters $\lambda_{i,f}$ and $\kappa_0$. 
In order to search for a protocol $(\kappa(t),\lambda(t))$ realizing the envisaged process, we plug the ansatz~\eqref{eq:pdfansatz} into the Fokker-Planck equation for the evolution of the pdf (see SM~\cite{SM} for details on the calculations). It is convenient to look for solutions with constant $\widetilde{\kappa}(t)=\kappa_0$; with this choice, one finds that the family of  solutions
\begin{subequations}
\label{eq:ev}
 \begin{eqnarray}
 \label{eq:betaev}
 \kappa(t) &= &\frac{\dot{\widetilde{\alpha}}(t)}{2\widetilde{\alpha}(t)}+ \kappa_0\widetilde{\alpha}(t)\\
 \label{eq:lambdaev}
\lambda(t)  &= &\lambda_i \exp \bbr{-\int_0^{t}[\kappa(t')- \kappa_0]dt'}\,\\
\widetilde{\kappa}(t)&= &\kappa_0\,\\
\widetilde{\lambda}(t)&=&\lambda(t)\,, 
\end{eqnarray} 
\end{subequations} satisfies the evolution equation. The function $\widetilde{\alpha}(t)$ appearing in Eqs.~\eqref{eq:ev} is arbitrary, among those that fulfill the boundary conditions~\eqref{eq:cond}; once it is chosen, the protocol is uniquely determined. The freedom on $\widetilde{\alpha}(t)$ provides a wide class of eligible SST for the process.
This explicit solution represents our main result.

In the solution worked out, $\widetilde{\kappa}=\kappa_0$ is constant during the whole process. This is a relevant simplification, because it implies that the coefficients $C_{n,l}^{(m)}(\widetilde{\kappa})$ appearing in the functional form of the pdf~\eqref{eq:statpdf} are also constant in time, and their derivatives do not appear in the calculations. By keeping $\widetilde{\kappa}$ fixed, we explore a two-dimensional manifold in the 3-dimensional space of pdf of the form~\eqref{eq:pdfansatz}: an even wider class of protocols may be searched for by allowing this parameter to vary in time, at the price of significantly more involved calculations. 


\textit{Controlled protocols --- } As alluded to in the Introduction, the degree of activity and the stiffness of the external potential can be controlled in actual experiments, within parameter ranges depending on the considered setup. In order to show that the analytical results of this Letter are in principle applicable to realistic experimental situations, it is useful to recall a couple of examples. With the setup described by Buttinoni et al.~\cite{buttinoni2012active}, spherical Janus particles with radius \textbf{$R=1 \, \mu m$} can have a persistent velocity varying in the interval $0 \,\mu m /s \le u_0 \le 1 \,\mu m /s$, depending on the intensity of the surrounding light. The rotational diffusivity has been measured to be $D_{\theta} \simeq 0.08 \,s^{-1}$. By calling  $\eta$ the dynamic viscosity of the fluid, $T$ the temperature and $k_B$ the Boltzmann constant, one gets the following equation for the translational diffusivity of the particles (not measured in the paper):
\begin{equation}
    D_t=\frac{k_B T}{6 \pi \eta R}=\frac{4}{3}R^2 D_{\theta}\simeq 0.10 \,\mu m^2/s\,,
\end{equation}
in agreement with the estimation provided in Ref.~\cite{takatori2016acoustic} for a similar situation. The dimensionless parameter $\lambda=u_0/ \sqrt{D_{\theta}D_t}$ can be thus tuned in the interval
\begin{equation}
    0\le \lambda \le 11\,.
\end{equation}
Slightly different results are found in Ref~\cite{buttinoni2013active}, corresponding to an even wider range for $\lambda$.
The particles may be confined in a quasi-harmonic, controllable potential as done in Ref.~\cite{takatori2016acoustic}, where acoustic waves are employed to trap a system of Janus particles with different chemical properties but similar radius. In that paper, two experimental situations are studied, in which particles with $D_{\theta}$ between $0.2 s^{-1}$ and $0.5 s^{-1}$ attain states with $\kappa=0.29$ and $\kappa=1.76$. 
Taking into account the different characteristic time for rotations, the dimensionless stiffness for the system described in Ref.~\cite{buttinoni2012active} can be expected to be tunable, at least, within the interval $1.2 \le \kappa \le 7$. A lower bound to the stiffness is expected to hold in experimental setups to prevent particles from moving away from the trap.

\begin{figure}
    \centering
    \includegraphics[width=0.99\linewidth]{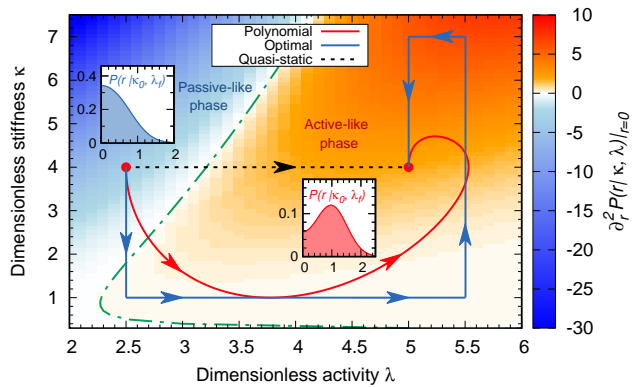}
    \caption{Parameter space of the model. The color code represents the value of $\partial_r^2P_s(r)|_{r=0}$, which is zero at the interface between the passive-like and the active-like phase (green dash-dotted curve). The black dotted line represents the path of a quasi-static protocol in which $\lambda$ is slowly varied between $\lambda_i=2.5$ and $\lambda_f=5$, while $\kappa=\kappa_0=4$ is kept constant. The red solid curve describes the solution to Eqs.~\eqref{eq:ev} associated to the polynomial protocol~\eqref{eq:poly}; in this case the final time $t_f$ is chosen in such a way that $\kappa(t)$ does not exceed the bounds $1 <\kappa< 7$, inspired by the experimental constraints discussed in the text. The blue lines show the minimal-time protocol, the dashed branches representing instantaneous change in the control parameter $\kappa$. Plots of the position pdf for the initial and the final states are also shown.}
    \label{fig:ps}
\end{figure}

In Fig.~\ref{fig:ps} the parameter space of the model is sketched. As in Ref.~\cite{Malakar2020Steady}, we distinguish between a passive-like phase characterized by $\partial_r^2P_s(0,\chi)<0$ and an active-like one where the particles tend to escape from the center of the potential and $\partial_r^2P_s(0,\chi)>0$. The range of the control parameters that is expected to be reached in experiments includes both passive-like and active-like stationary distributions, and it is interesting to search for SST between these two states.

Possibly, the simplest way to find an explicit smooth protocol satisfying Eqs.~\eqref{eq:ev} is to enforce a polynomial form for $\widetilde{\alpha}(t)$. We have to impose the boundary conditions Eq.~\eqref{eq:condalpha} and the final condition for $\lambda$. If we also require that 
\begin{equation}
\label{eq:betaacond}
    \kappa(0)=\kappa(t_f)=\kappa_0\,,
\end{equation}
i.e. that the stiffness be varied continuously without abrupt changes at the beginning and at the end of the protocol, five degrees of freedom are needed. The polynomial needs therefore to be at least fourth order, i.e.,
\begin{equation}
\label{eq:poly}
\begin{aligned}
\widetilde{\alpha}(t)&=1+\sum_{n=2}^4 \widetilde{\alpha}_n t^n \quad \text{with}\\
 \widetilde{\alpha}_2=-\frac{30 }{\kappa_0 t_f^3}\ln \frac{\lambda_f}{\lambda_i}&,\quad \widetilde{\alpha}_3=-\frac{2 \widetilde{\alpha}_2}{t_f},\quad \widetilde{\alpha}_4=\frac{\widetilde{\alpha}_2}{t_f^2}\,.
 \end{aligned}
\end{equation}
In Fig.~\ref{fig:ps}, the red solid curve shows a protocol of this sort for a realistic situation, bringing the state of the system from the passive- to the active-like phase in a time $\tau\simeq 0.66 D_{\theta}^{-1}$. 
Spontaneous relaxation to the stationary state is expected to occur for $\tau > \tau_r$, where $\tau_r= D_{\theta}^{-1}$ is the typical time-scale related to the rotational motion~\cite{basu2018}.

\textit{Minimal time --- } As discussed before, experimental conditions often impose bounds of the kind 
\begin{equation}
\label{eq:bounds}
    \kappa_{-}\le \kappa(t)\le \kappa_{+}
\end{equation} on the enforceable stiffness.  Our interest now goes to finding the fastest protocol subjected to such a constraint (i.e., the one with the shortest 
connecting time $t_f^{min}$), among all those encoded in the form \eqref{eq:pdfansatz}. This amounts to 
identifying the optimal function $\widetilde{\alpha}(t)$, from which the driving parameters $\kappa(t)$ and $\lambda(t)$ follow.

\begin{figure}
    \centering
    \includegraphics[width=0.99\linewidth]{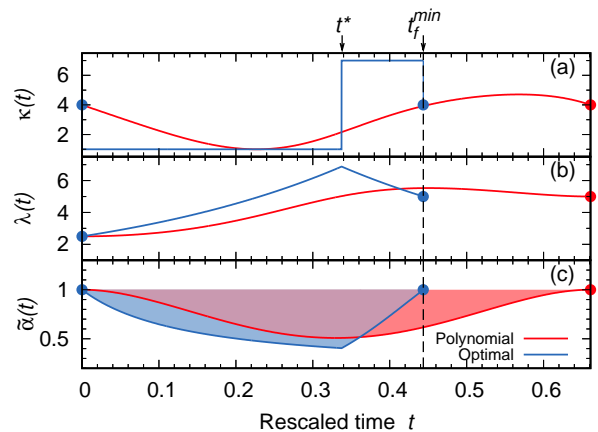}
    \caption{Evolution of the parameters for the minimum-time protocol and for the solution to Eqs.~\eqref{eq:ev} associated to the polynomial with coefficients~\eqref{eq:poly}. Panel (a) shows the time dependence of $\kappa(t)$, panel (b) that of $\lambda(t)$ and panel (c) the evolution of $\widetilde{\alpha}(t)$. The dashed vertical line identifies the minimum time over all possible protocols of the type given by Eq.~\eqref{eq:pdfansatz}, with $1<\kappa<7$. The switching time $t^{\star}$ is also highlighted on the top axis.
    The shaded areas in panel (c) do not depend on the protocol, once the r.h.s. of Eq.~\eqref{eq:area} is fixed (here $\kappa_0=4$, $\lambda_i=2.5$ and $\lambda_f=5$).}
    \label{fig:td}
\end{figure}

\begin{figure}
    \centering
    \includegraphics[width=0.99\linewidth]{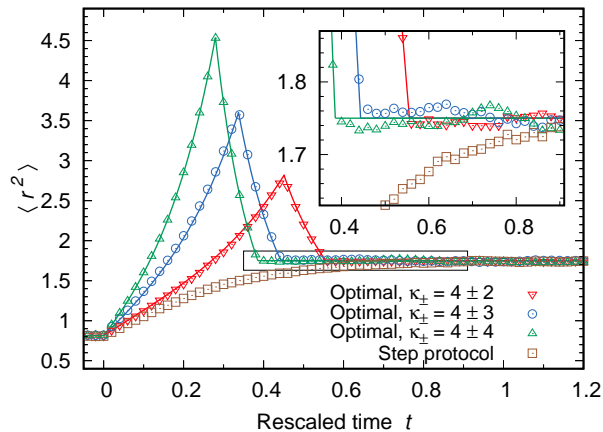}
    \caption{Comparison between the relaxation dynamics induced by a step-protocol and the optimal control protocol. In the former case (brown squares), the value of $\lambda$ is suddenly changed from $\lambda_i$ to $\lambda_f$ at $t=0$; in the latter the optimal protocol described in the text is operated, for different values of the bounds $\kappa_{\pm}$. Symbols refer to the instantaneous value of $\av{r^2}$, computed as an average over $10^4$ independent numerical simulations (see SM~\cite{SM} for details). Solid lines provide the analytical prediction (only for the controlled protocols). The inset is a zoom on the area enclosed by the rectangle.}
    \label{fig:comp}
\end{figure}

We will consider the case in which the activity of the particles is increased during the process, the reverse case being analogous.
It is useful to note that, plugging Eq.~\eqref{eq:betaev} into Eq.~\eqref{eq:lambdaev}, one has
\begin{equation}
\label{eq:area}
    \frac{1}{\kappa_0}\ln\frac{\lambda_f}{\lambda_0}=\int_0^{t_f}dt[1-\widetilde{\alpha}(t)]\,,
\end{equation}
i.e. the area between $\widetilde{\alpha}(t)$ and the line $\widetilde{\alpha}=1$ is determined, once $\lambda_0$, $\lambda_f$ and $\kappa_0$ are fixed, and it does not depend on $t_f$. Minimizing the integration interval $t_f$ in the r.h.s. of Eq.~\eqref{eq:area}, once the l.h.s. is fixed, is thus equivalent to maximizing the integrand. Taking into account the boundary conditions~\eqref{eq:condalpha}, the minimal $t_f$ is therefore obtained by first decreasing $\widetilde{\alpha}(t)$ as quickly as possible, and then bringing it back to $1$, again as quickly as the bounds on $\kappa$ allow, in such a way that Eq.~\eqref{eq:area} is verified. The conditions~\eqref{eq:bounds}, using Eq.~\eqref{eq:betaev}, imply
 \begin{equation}
\label{eq:constrbeta}
    2 \widetilde{\alpha}(t)[\kappa_{-}- \kappa_0 \widetilde{\alpha}(t)]  \le \dot{\widetilde{\alpha}}(t)\le 2 \widetilde{\alpha}(t)[\kappa_{+}- \kappa_0 \widetilde{\alpha}(t)] \,.
 \end{equation}
The two limiting curves $\widetilde{\alpha}_{-}(t)$ and $\widetilde{\alpha}_{+}(t)$ (obtained by imposing the least and the largest value of $\dot{\widetilde{\alpha}}(t)$, respectively) are thus:
\begin{eqnarray}
\widetilde{\alpha}_{-}(t)=&\kappa_-\sbr{\kappa_0-(\kappa_0-\kappa_-)e^{-2\kappa_- t}}^{-1}\\
\widetilde{\alpha}_{+}(t)=&{\kappa_+}\sbr{\kappa_0-(\kappa_0-\kappa_+)e^{2\kappa_+(t_f^{min}-t) }}^{-1}\,,
\end{eqnarray}
where the boundary conditions~\eqref{eq:condalpha} have been enforced. In the light of the above considerations, we need to alternate a maximal decompression ($\widetilde{\alpha}(t)=\widetilde{\alpha}_-(t)$) and a maximal compression ($\widetilde{\alpha}(t)=\widetilde{\alpha}_{+}(t)$). This class of protocols is usually encountered when minimizing the duration of linear processes; they are referred to as ``bang-bang protocols''~\cite{kirk2004optimal,PhysRevResearch.3.023128}. Since they involve unphysical discontinuities in the control parameters, they have to be understood as limits of continuous protocols that are actually realizable in practice. In the SM~\cite{SM} we explore this aspect in some detail. Let us denote by $t^{\star}$ the time at which the two regimes are switched. The continuity condition on $\widetilde{\alpha}(t)$ yields
\begin{equation}
    \widetilde{\alpha}_{-}(t^{\star})=\widetilde{\alpha}_{+}(t^{\star})\equiv \widetilde{\alpha}^{\star}\,,
\end{equation}
while from Eq.~\eqref{eq:area} one obtains, by integration,
\begin{equation}
    \ln\cbr{\frac{\lambda_f}{\widetilde{\alpha}^{\star} \lambda_i}}=\kappa_0 t_f^{min} -\kappa_- t^{\star}-\kappa_+ (t_f^{min}-t^{\star})\,.
\end{equation}
The above equations can be solved numerically for $t^{\star}$ and $t_f^{min}$ (see SM~\cite{SM} for a plot of $t_f^{min}$ as a function of the boundary conditions). In Figure~\ref{fig:ps}, the blue curve represents the optimal protocol in the parameter space under some realistic constraints. The time dependence of the parameters is presented in Fig.~\ref{fig:td}, where also the smooth polynomial protocol discussed before is shown for comparison. In panel~\ref{fig:td}(c) the equivalence of the areas discussed above can be appreciated for the two considered processes. Figure~\ref{fig:comp} shows a comparison with the relaxation induced by a step-protocol in which $\lambda$ is suddenly switched at $t=0$ from $\lambda_i$ to $\lambda_f$. Here we consider the dynamics of the observable $\av{r^2}$, the variance of the radial position (the average is computed over many realizations of the protocol). Details on the analytical form of the observable, as well as on the numerical simulations performed, can be found in the SM~\cite{SM}. Within the already existing experimental conditions described before, by using the proposed optimal protocol it is possible to decrease the duration of the process by a factor 2.

Here we have assumed that the stiffness of the confining potential can be varied discontinuously. In the SM~\cite{SM} we show that this limit protocol can be approached with arbitrary precision by continuous-in-time protocols. 

The minimal time $t_f^{min}$ should not be interpreted as a definitive bound, as it has been derived by only considering solutions of the form~\eqref{eq:pdfansatz} with constant $\widetilde{\kappa}=\kappa_0$: even faster protocols might be achievable, in principle, by allowing for more general functional forms of  our ansatz. 
 
\textit{Conclusions --- } In the present letter, we have discussed a class of exact analytical protocols to bring an ABP system from an initial non-equilibrium stationary state to another final stationary state having a different degree of activity, in a given time.
Among this family of protocols, we have also identified the one leading to the minimal time. The proposed protocols are expected to be relevant in actual experiments with tunable active particles, 

The present work extends the quest for controlling stochastic motion to the realm of active particles. To the best of our knowledge, this is the first case in which SST can be found for this class of systems, and one of the few involving out-of-equilibrium models~\cite{baldassarri2020engineered,PhysRevResearch.3.023128,baldovin2022shortcuts}.

Our computation is the starting point for the solution of other optimal problems for ABPs: for instance, the average work done during a realization can be computed~\cite{SM} and minimized with analytical methods, a task that has been so far accomplished, for active models, only with numerical techniques~\cite{PhysRevE.99.022605}.
Since 
our search for the optimal protocol is restricted to the class of processes fulfilling condition~\eqref{eq:pdfansatz}, a further step would consist in proving (or excluding) that the ``global'' optimum belongs to this family, making use of Pontryagin's principle~\cite{pontryagin1987mathematical}.
 Protocols connecting states with different stiffness may be also searched for, following similar approaches. 
Future developments pertain to
the search for SSTs in three dimensions~\cite{turci2021phase} (e.g., in the presence of homogeneous external force~\cite{vachier2019dynamics}), and for 
interacting particles~\cite{pototsky2012active, solon2015pressure}; the latter has been studied in the context of passive systems~\cite{salambo2020engineered}, but with few degrees of freedom only.

 Similar strategies may be attempted for active particle models whose stationary state is analytically known, as the 1D Run-and-Tumble~\cite{tailleur2009sedimentation,PhysRevE.99.032132} or the Active Ornstein-Uhlenbeck particles with Unified Color Noise approximation~\cite{maggi2015multidimensional,caprini2019activity}. 
We emphasize that our method, based on suitable deformations of the stationary distribution, may be used to search for SST 
in different contexts, provided that the stationary state is known. Finding the general conditions to be fulfilled for this approach to provide a suitable solution is 
an interesting research perspective.


\acknowledgements{The authors thankfully acknowledge useful discussions with P. Bayati, L. Caprini and A. Puglisi.}

\bibliography{biblio}

\end{document}


\title{Supplemental Material for ``Control of Active Brownian Particles: an exact solution''}

\author{Marco Baldovin}
\affiliation{Universit\'{e}  Paris-Saclay,  CNRS,  LPTMS,  91405,  Orsay,  France}%
 \affiliation{Institute for Complex Systems, CNR, 00185, Rome, Italy}
 \email{marco.baldovin@cnr.it}
\author{David Gu\'{e}ry-Odelin}%
\affiliation{Laboratoire  Collisions,  Agr\'{e}egats,  R\'{e}eactivit\'{e}e,  FeRMI,  Universit\'{e}  de  Toulouse,  CNRS,  UPS,  France}%
\author{Emmanuel Trizac}%
\affiliation{Universit\'{e}  Paris-Saclay,  CNRS,  LPTMS,  91405,  Orsay,  France}
\affiliation{Ecole normale sup\'erieure de Lyon, F-69342 Lyon, France}%

\date{\today}

\maketitle

\textbf{Structure of the Supplemental Material:} In Sections~\ref{sec:sm_defc} and~\ref{sec:appendixlaguerre} we recall the stationary solution for the 2-dimensional Active Brownian Particle model in a harmonic potential, derived in Ref.~\cite{Malakar2020Steady}, and the properties of the Generalized Laguerre polynomials, which will turn out to be useful for the following calculations. In Section~\ref{sec:sm_derivation}, Eqs.~(\eqmain) is derived. This is the main result of the paper. Section~\ref{sec:appw} is devoted to the calculation of the average work and of the observable $\av{r^2}$. Section~\ref{sec:sm_minimaltime} contains a detailed characterization of the minimal-time protocol: first the minimal time is studied numerically as a function of the system's parameters; it is then shown how the optimal protocol is approached by considering high-order polynomials; finally an approximation with continuous-in-time functions is discussed. We present in Section~\ref{sec:simul} some details on the direct numerical integration of the Langevin equation.

\section{Stationary solution}
\label{sec:sm_defc}

The stationary solution for a 2-dimensional Active Brownian Particle in a harmonic potential has been described in Ref.~\cite{Malakar2020Steady}. Here we briefly recall the result for completeness, as it will be useful for the following.
 It is convenient to recast the problem in terms of a Fokker-Planck equation for the evolution of the probability density function:
\begin{equation}
\label{eq:fp0}
 \partial_t P(\vect{r},\theta) = - \vect{\nabla}\bbr{\sbr{\lambda \widehat{\vect{e}}(\theta)-\kappa \vect{r}-\frac{1}{\alpha}\vect{\nabla} }P(\vect{r},\theta)} +\partial_{\theta}^2 P(\vect{r},\theta)\,,
\end{equation}
where the dimensionless parameters have been introduced in Eqs.~(2) of the main text. Let us call $\varphi$ the angle between the $x$ axis of the plane where the particle moves and the vector $\vect{r}$. Assuming isotropy, the density is expected to depend only on the difference $\chi=\theta-\varphi$. Eq.~\eqref{eq:fp0} can be thus recast as
\begin{equation}
\label{eq:fp}
  \partial_t P=\mathcal{L}_r P-\lambda \cos\chi\partial_r P+\lambda \sin\chi\frac{1}{r}\partial_{\chi} P+\frac{1}{\alpha r^2}\partial^2_{\chi} P+\partial^2_{\chi} P
\end{equation} 
where
\begin{equation}
 \mathcal{L}_r f(r)=\frac{1}{r}\partial_r\sbr{r\cbr{\frac{1}{\alpha}\partial_r+\kappa r}f(r)}\,.
\end{equation}

It can be verified that the stationary solution for the above equation is

\begin{equation}
\label{eq:statpdf}
 P_{s}(r,\chi)=F(r,\chi|\kappa, \lambda,\alpha)\equiv\alpha\sum_{m=0}^{\infty}\lambda^m \alpha^{m/2}\sum_{2n+|l|=m}C_{n,l}^{(m)}(\kappa)\phi_{n,l}(\sqrt{\alpha}r,\chi|\kappa)\,,
\end{equation}
where the functions $\phi_{n,l}(r,\chi)$ are defined by
\begin{equation}
\label{eq:phi}
 \phi_{n,l}(r,\chi|\kappa)=\sbr{\frac{n! \cbr{\frac{\kappa}{2}}^{|l|+1}}{\pi (n + |l|)!}}^{\frac{1}{2}}r^{|l|}e^{-\frac{\kappa r^2}{2}}L^{|l|}_n\cbr{\frac{\kappa r^2}{2}}e^{il\chi}\,,
\end{equation} 
$L^{l}_n(x)$ being the generalized Laguerre polynomials (see Section~\ref{sec:appendixlaguerre} of this SM for their definition and properties). The second sum in Eq.~\eqref{eq:statpdf} is constrained to the (integer) values of $l$ and $n\ge 0$ such that $2n+|l|=m$. The coefficients $C_{n,l}^{(m)}(\kappa)$ are defined recursively, according to the following iterative rules (the $\kappa$ dependence is dropped to avoid clutter):
\begin{subequations}
 \begin{eqnarray}
 \label{eq:normalc000}
 C_{0,0}^{(0)}&=&\sqrt{\frac{\kappa}{2 \pi}}\\
 \label{eq:normalc0ll}
 C_{0,l}^{(l)}&=&\frac{C_{0,l-1}^{(l-1)} \sqrt{l\frac{\kappa}{2}}  }{ \kappa l + l^2  }\quad\quad l>0\\
 \label{eq:normalcn02n}
 C_{n,0}^{(2n)}&=&-\frac{   C_{n-1,1}^{(2n-1)}     }{ \sqrt{2\kappa n}  }\quad\quad n>0\\
\label{eq:iterc}
 C_{n,l}^{(m)}&=&\frac{C_{n,l-1}^{(m-1)} \sqrt{(n+l)\frac{\kappa}{2}}  -   C_{n-1,l+1}^{(m-1)}  \sqrt{n\frac{\kappa}{2}}    }{ \kappa (2n+l) + l^2  }\quad\quad l>0, n>0\,. 
\end{eqnarray} 
\end{subequations}
The coefficients with negative values of $l$ are defined by the symmetry relation
\begin{equation}
\label{eq:sym}
    C_{n,-l}^{(m)}=C_{n,l}^{(m)}\,.
\end{equation}

With the above prescriptions, one can first compute $C_{0,l}^{(l)}$ iteratively for every value of $l>0$, then all terms $C_{1,l}^{l+1}$, $C_{2,l}^{l+2}$, and so on, and finally use the symmetry relation~\eqref{eq:sym} to determine the remaining coefficients.

Let us verify that the proposed solution is properly normalized for any choice of the parameters, i.e. that the quantity
\begin{equation}
 \int_0^{2\pi}d\chi \int_0^{\infty} dr\, r F(r,\chi|\kappa, \lambda, \alpha)=\int_0^{2\pi}d\chi \int_0^{\infty} dr\, r \sum_{m=0}^{\infty}\lambda^m \alpha^{\frac{m}{2}+1} \sum_{2n+|l|=m}C_{n,l}^{(m)}(\kappa)\phi_{n,l}(\sqrt{\alpha}r\chi|\kappa)
\end{equation}
is equal to 1. 
First of all, let us notice that all terms with $l \ne 0$ are null, due to the factor $e^{i l \chi}$ which turns into a Kronecker delta $2 \pi \delta_{l,0}$ once it is integrated over $\chi$. The second sum is constrained to $2n+|l|=m$, which implies therefore $m=2n$ for all nonvanishing terms. We can thus write
\begin{equation}
\begin{aligned}
 \int_0^{2\pi}d\chi \int_0^{\infty} dr\, r F(r,\chi|\kappa,\lambda, \alpha)=&\sqrt{2\pi \kappa} \int_0^{\infty} dr\, r \sum_{n=0}^{\infty}\lambda^{2n} \alpha^{n+1} C_{n,0}^{(2n)}(\kappa)e^{-\frac{\alpha \kappa r^2}{2}}L^{0}_n\cbr{\frac{\alpha \kappa r^2}{2}}\\
 =&\sqrt{\frac{2\pi}{ \kappa}} \sum_{n=0}^{\infty}\lambda^{2n}\alpha^{n} C_{n,0}^{(2n)}(\kappa)\int_0^{\infty} du\, e^{-u}L^{0}_n\cbr{u}\\
 =&\sqrt{\frac{2\pi}{ \kappa}} \sum_{n=0}^{\infty}\lambda^{2n}\alpha^{n} C_{n,0}^{(2n)}(\kappa)\frac{1}{n!}\int_0^{\infty} du\, \frac{d^{n}}{du^n}\cbr{e^{-u}u^n}\,,
\end{aligned}
\end{equation}
where we first implemented the variable change $u=\alpha \kappa r^2/2$ and then, in the last step, we exploited the Rodrigues formula for Laguerre polynomials (see Eq.~\eqref{eq:rodrigues} of Appendix~\ref{sec:appendixlaguerre}). The integral is only different from zero (and equal to 1) when $n=0$, so that one finally has:
\begin{equation}
 \int_0^{2\pi}d\chi \int_0^{\infty} dr\, r F(r,\chi|\kappa,\lambda, \alpha)=\sqrt{\frac{2\pi}{ \kappa}}  C_{0,0}^{(0)}(\kappa)=1\,.
\end{equation}

\section{Some properties of the Generalized Laguerre Polynomials}
\label{sec:appendixlaguerre}
The Generalized Laguerre polynomials are defined as
\begin{equation}
\label{eq:lagdef}
 L_n^{(y)}(x)=\sum_{i=0}^n(-1)^i \binom{n+y}{n-i}\frac{x^i}{i!}\,,
\end{equation}
or, alternatively, through the Rodrigues formula
\begin{equation}
\label{eq:rodrigues}
  L_n^{(y)}(x)=\frac{x^{-y} e^x}{n!}\frac{d^n}{dx^n}\cbr{e^{-x}x^{n+y}}\,.
\end{equation} 
These functions generalize the Laguerre polynomials (which can be recovered by setting $y=0$).

It can be shown that, with the above definition, the derivative of a generalized Laguerre polynomial assumes the form
\begin{equation}
\label{eq:lagder}
 \frac{d^k}{dx^k}L_n^{(y)}(x)=\begin{cases}
  (-1)^kL_{n-k}^{(y+k)}(x)\quad &\text{if}\quad k\le n\\
  0\quad\quad &\text{otherwise}\,.
 \end{cases}
\end{equation} 
Several recurrence formulas can be defined for this set of functions. Two particularly useful ones are the following:
\begin{equation}
\label{eq:rec1}
 L_{n}^{(y)}(x)=L_{n}^{(y+1)}(x)-L_{n-1}^{(y+1)}(x)
\end{equation} 
\begin{equation}
\label{eq:rec2}
L_n^{(y)}(x)=\frac{y + 1-x}{n}L_{n-1}^{(y+1)}(x)-\frac{x}{n}L_{n-2}^{(y+2)}(x)\,.
\end{equation}

\section{Derivation of Eqs.~(\eqmain)}
\label{sec:sm_derivation}
To derive Eqs.~(\eqmain) of the main text, one has to plug the ansatz
\begin{equation}
\label{eq:pdfansatz}
    P(r,\chi)=F(r,\chi|\tilde{\kappa},\tilde{\lambda}, \tilde{\alpha})=\sum_{m=0}^{\infty}\tilde{\lambda}^m \tilde{\alpha}^{\frac{m}{2}+1} \sum_{2n+|l|=m}C_{n,l}^{(m)}(\tilde{\kappa})\phi_{n,l}(\sqrt{\tilde{\alpha}}r,\chi|\tilde{\kappa})\,,
\end{equation}
with $\phi_{n,l}$ defined by Eq.~\eqref{eq:phi}, into the Fokker-Planck equation
\begin{equation}
    \partial_t P=\mathcal{L}_r P-\lambda \cos\chi\partial_r P+\lambda \sin\chi\frac{1}{r}\partial_{\chi} P+\frac{1}{r^2}\partial^2_{\chi} P+\partial^2_{\chi} P\,;
\end{equation}
Imposing the validity of the evolution equation for all values of $r$ and $\chi$, one aims at deriving a set of relations for the controlling parameters $\kappa$ and $\lambda$, as well as for the parameters of the pdf $\tilde{\kappa}$, $\tilde{\lambda}$ and $\tilde{\alpha}$.

We search for solutions that satisfy
\begin{equation}
\label{eq:condition}
    \tilde{\kappa}(t)=\kappa_0\,;
\end{equation}
as it will soon be clear, this additional condition leads to a remarkable simplification of the calculations, as it implies that the coefficients $C_{n,l}^{(m)}(\tilde{\kappa})$ are also constant in time. It will consequently not be necessary to compute the (quite involved) derivatives of these recursively-defined coefficients. 

For the sake of readability, we will henceforth drop explicit dependence on the parameters and consider
$$
P\equiv P(r,\chi)\quad C_{n,l}^{(m)}\equiv C_{n,l}^{(m)}(\kappa_0) \quad\phi_{n,l}\equiv \phi_{n,l}(\sqrt{\tilde{\alpha}(t)}r,\chi|\kappa_0)\quad L_{n}^{l}\equiv L_{n}^{l}\cbr{\frac{\tilde{\alpha}(t)\kappa_0 r^2}{2}}\,.
$$

We start with the time derivative $\partial_t P$. Our ansatz only depends on time through $\tilde{\lambda}(t)$ and $\tilde{\alpha}(t)$. One obtains
\begin{equation}
\label{eq:termt}
\partial_tP=\sum_{m=0}^{\infty}\tilde{\lambda}^{m}\sqrt{\tilde{\alpha}^{m+2}}\sum_{2n+|l|=m}C_{n,l}^{(m)}\sbr{m\cbr{\frac{\dot{\tilde{\lambda}}}{\tilde{\lambda}} +\frac{\dot{\tilde{\alpha}}}{2\tilde{\alpha}}}+\frac{\dot{\tilde{\alpha}}}{2\tilde{\alpha}}\cbr{|l|+2-\tilde{\alpha}\kappa r^2-\tilde{\alpha}\kappa r^2\frac{L_{n-1}^{|l|+1}}{L_{n}^{|l|}}}}\phi_{n,l}\,,
\end{equation}
where the rule for the derivative of generalized Laguerre polynomials, Eq.~\eqref{eq:lagder}, has been exploited. Please notice that without condition~\eqref{eq:condition}, the above term would include an additional contribution proportional to $\dot{\tilde{\kappa}}$, where the derivatives of $C_{n,l}^{(m)}$ would be also present.

The term $\mathcal{L}_rP$ is more involved, and its evaluation requires the 
use of the recursion formulas for the generalized Laguerre Polynomials. It is important to note that inside the operator $\mathcal{L}_r$ one has the time-dependent parameter $\kappa$ (which determines the stiffness of the external potential).
\begin{equation}
\begin{aligned}
\label{eq:termr}
\mathcal{L}_rP= \sum_{m=0}^{\infty}\tilde{\lambda}^{m}\sqrt{\tilde{\alpha}^{m+2}}&\sum_{2n+|l|=m}C_{n,l}^{(m)}\frac{1}{r}\frac{\partial}{\partial r}\sbr{r\cbr{\frac{\partial \phi_{n,l}}{\partial r}+\kappa r \phi_{n,l}}}\\
= \sum_{m=0}^{\infty}\tilde{\lambda}^{m}\sqrt{\tilde{\alpha}^{m+2}}&\sum_{2n+|l|=m}C_{n,l}^{(m)}\frac{1}{r}\frac{\partial}{\partial r}\sbr{\cbr{|l|-\tilde{\alpha}\kappa_0 r^2 \frac{L_{n-1}^{|l|+1}}{L_{n}^{|l|}}+(\kappa-\tilde{\alpha} \kappa_0) r}\phi_{n,l}}\\
= \sum_{m=0}^{\infty}\tilde{\lambda}^{m}\sqrt{\tilde{\alpha}^{m+2}}&\sum_{2n+|l|=m}C_{n,l}^{(m)}\cbr{\frac{l^2}{r^2}-\tilde{\alpha}\kappa_0|l|+\sbr{-2(|l|+2)\tilde{\alpha}\kappa_0  +\tilde{\alpha}^2\kappa_0^2r^2} \frac{L_{n-1}^{|l|+1}}{L_{n}^{|l|}} +\tilde{\alpha}^2\kappa_0^2r^2 \frac{L_{n-2}^{|l|+1}}{L_{n}^{|l|}} }\phi_{n,l} +\\
&+\sum_{m=0}^{\infty}\tilde{\lambda}^{m}\sqrt{\tilde{\alpha}^{m+2}}\sum_{2n+|l|=m}C_{n,l}^{(m)}(\kappa-\tilde{\alpha} \kappa_0)\cbr{ |l|+2-\tilde{\alpha} \kappa_0r^2 - \tilde{\alpha} \kappa_0r^2\frac{L_{n-1}^{|l|+1}}{L_{n}^{|l|}} }\phi_{n,l}\\
=\sum_{m=0}^{\infty}\tilde{\lambda}^{m}\sqrt{\tilde{\alpha}^{m+2}}&\sum_{2n+|l|=m}C_{n,l}^{(m)}\sbr{\frac{l^2}{r^2}-\tilde{\alpha}\kappa_0|l|-2n\tilde{\alpha}\kappa_0+(\kappa-\tilde{\alpha} \kappa_0)\cbr{|l|+2-\tilde{\alpha}\kappa_0 r^2-\tilde{\alpha}\kappa_0 r^2 \frac{L_{n-1}^{|l|+1}}{L_{n}^{|l|}}}}\phi_{n,l}\\
=\sum_{m=0}^{\infty}\tilde{\lambda}^{m}\sqrt{\tilde{\alpha}^{m+2}}&\sum_{2n+|l|=m}C_{n,l}^{(m)}\sbr{\frac{l^2}{r^2}-m\tilde{\alpha}\kappa_0+(\kappa-\tilde{\alpha} \kappa_0)\cbr{|l|+2-\tilde{\alpha}\kappa_0 r^2-\tilde{\alpha}\kappa_0 r^2 \frac{L_{n-1}^{|l|+1}}{L_{n}^{|l|}}}}\phi_{n,l}
\end{aligned}
\end{equation} 
where, in the last step, we made use of Eq.~\eqref{eq:rec2}.

Since the density $P$ only depends on $\chi$ through the factor $e^{i l\chi }$ appearing in the definition of $\phi$, the terms of the Fokker-Planck equation which only involve the second derivative with respect to $\chi$ read
\begin{equation}
\label{eq:termchi2}
 \frac{1}{r^2}\frac{\partial^2 P}{\partial \chi^2}+\frac{\partial^2 P}{\partial \chi^2}=\sum_{m=0}^{\infty}\tilde{\lambda}^{m}\sqrt{\tilde{\alpha}^{m+2}}\sum_{2n+|l|=m}C_{n,l}^{(m)}\sbr{-\frac{l^2}{r^2}-l^2}\phi_{n,l}\,.
\end{equation} 

Finally, we have to evaluate the $\lambda$-dependent terms. First, we have that
\begin{equation}
 -\lambda \cos \chi \partial_r P =-\cos\chi\sum_{m=0}^{\infty}\lambda\tilde{\lambda}^{m}\sqrt{\tilde{\alpha}^{m+2}}\sum_{2n+|l|=m}C_{n,l}^{(m)}\cbr{|l|r^{-1}-\tilde{\alpha}\kappa_0 r-\tilde{\alpha}\kappa_0 r \frac{L_{n-1}^{|l|+1}}{L_{n}^{|l|}}}\phi_{n,l}\,.
\end{equation} 
Here we have to apply in sequence Eqs.~\eqref{eq:rec1} and~\eqref{eq:rec2}, leading to
\begin{equation}
\label{eq:stepcos}
\begin{aligned}
 &-\lambda \cos \chi \partial_r P\\=&-\cos\chi\sum_{m=0}^{\infty}\lambda\tilde{\lambda}^{m}\sqrt{\tilde{\alpha}^{m+2}}\sum_{2n+|l|=m}C_{n,l}^{(m)}\cbr{(n+1)r^{-1}\frac{L_{n+1}^{|l|-1}}{L_{n}^{|l|}}-\frac{\tilde{\alpha}\kappa_0r}{2} \frac{L_{n}^{|l|+1}}{L_{n}^{|l|}}}\phi_{n,l}\\
 =&-\sum_{n=0}^{\infty}\sum_{|l|>0}\delta_{m,2n+|l|}\lambda\tilde{\lambda}^{m}\sqrt{\tilde{\alpha}^{m+2}}C_{n,l}^{(m)}\cos\chi   \cbr{(n+1)r^{-1}\frac{L_{n+1}^{|l|-1}}{L_{n}^{|l|}}-\frac{\tilde{\alpha}\kappa_0r}{2} \frac{L_{n}^{|l|+1}}{L_{n}^{|l|}}}\phi_{n,l}\\
 &-\sum_{n=0}^{\infty}\lambda\tilde{\lambda}^{2n}\tilde{\alpha}^{n+1}C_{n,l}^{(2n)}\cos\chi \sqrt{\frac{\kappa_0}{2\pi }} \cbr{(n+1)r^{-1}L_{n+1}^{-1}-\frac{\tilde{\alpha}\kappa_0r}{2} L_{n}^{1}}e^{-\frac{\tilde{\alpha} \kappa_0 r^2}{2}}\\
 =&-\sum_{n=0}^{\infty}\sum_{|l|>0}\delta_{m,2n+|l|}\lambda\tilde{\lambda}^{m}\sqrt{\tilde{\alpha}^{m+2}}C_{n,l}^{(m)}\cos\chi   \cbr{(n+1)r^{-1}\frac{L_{n+1}^{|l|-1}}{L_{n}^{|l|}}-\frac{\tilde{\alpha}\kappa_0r}{2} \frac{L_{n}^{|l|+1}}{L_{n}^{|l|}}}\phi_{n,l}\\
 &+2\sum_{n=0}^{\infty}\lambda\tilde{\lambda}^{2n}\tilde{\alpha}^{n+1}C_{n,l}^{(2n)}\cbr{\frac{e^{i\chi}}{2}+\frac{e^{-i\chi}}{2}} \sqrt{\frac{\kappa_0}{2\pi }} \frac{\tilde{\alpha}\kappa_0r}{2} L_{n}^{1}e^{-\frac{\tilde{\alpha} \kappa_0 r^2}{2}}\,.\\
\end{aligned}
\end{equation} 
 In the last step, we made use of the relation
 \begin{equation}
     (n+1)r^{-1}L_{n+1}^{-1}=-\frac{\tilde{\alpha} \kappa_0 r}{2}L_{n}^{0}-\frac{\tilde{\alpha} \kappa_0 r}{2}L_{n-1}^{1}=-\frac{\tilde{\alpha} \kappa_0 r}{2}L_{n}^{1}
 \end{equation}
 following from Eqs.~\eqref{eq:rec1} and~\eqref{eq:rec2}.

Next, we have to evaluate
\begin{equation}
\label{eq:stepsin}
 \begin{aligned}
  &\lambda \frac{\sin \chi}{r} \partial_{\chi} P =\sum_{m=0}^{\infty}\lambda\tilde{\lambda}^{m}\sqrt{\tilde{\alpha}^{m+2}}\sum_{2n+|l|=m}C_{n,l}^{(m)}\sin\chi il\phi_{n,l}\\
  &=\sum_{n=0}^{\infty}\sum_{|l|>0}\delta_{m,2n+|l|}\lambda\tilde{\lambda}^{m}\sqrt{\tilde{\alpha}^{m+2}}C_{n,l}^{(m)} \frac{il}{|l|}\sin\chi\cbr{(n+1)r^{-1}\frac{L_{n+1}^{|l|-1}}{L_{n}^{|l|}}+\frac{\tilde{\alpha} \kappa_0 r}{2} +\frac{\tilde{\alpha} \kappa_0 r}{2}\frac{L_{n-1}^{|l|+1}}{L_{n}^{|l|}}   }\phi_{n,l}\\
   &=\sum_{n=0}^{\infty}\sum_{|l|>0}\delta_{m,2n+|l|}\lambda\tilde{\lambda}^{m}\sqrt{\tilde{\alpha}^{m+2}}C_{n,l}^{(m)}\frac{il}{|l|}\sin\chi \cbr{(n+1)r^{-1}\frac{L_{n+1}^{|l|-1}}{L_{n}^{|l|}}-\frac{\tilde{\alpha}\kappa_0r}{2} \frac{L_{n}^{|l|+1}}{L_{n}^{|l|}}}\phi_{n,l} \,,
 \end{aligned}
\end{equation} 
where we first applied the recurrence formula~\eqref{eq:rec2} and then Eq.~\eqref{eq:rec1}. Putting together Eq.~\eqref{eq:stepcos} and~\eqref{eq:stepsin} and noticing that
\begin{equation}
 \cos \chi \pm \frac{il}{|l|}\sin \chi=e^{\pm i\frac{l}{|l|}\chi}
\end{equation} 
one gets
\begin{equation}
\label{eq:sincos1}
\begin{aligned}
 &-\lambda \cos \chi \partial_r P+\lambda \frac{\sin \chi}{r} \partial_{\chi} P=\\
 =&-\sum_{n=0}^{\infty}\sum_{|l|>0}\delta_{m,2n+|l|}\lambda\tilde{\lambda}^{m}\sqrt{\tilde{\alpha}^{m+2}}C_{n,l}^{(m)} e^{il \chi-i\frac{l}{|l|}\chi} \sbr{\frac{n! \cbr{\frac{ \kappa_0}{2}}^{|l|+1}}{\pi (n + |l|)!}}^{\frac{1}{2}} \sqrt{\tilde{\alpha}^{|l|}}(n+1)r^{|l|-1}L_{n+1}^{|l|-1}e^{-\frac{\tilde{\alpha} \kappa_0 r^2}{2}} \\
 &+\sum_{n=0}^{\infty}\sum_{|l|>0}\delta_{m,2n+|l|}\lambda\tilde{\lambda}^{m}\sqrt{\tilde{\alpha}^{m+2}}C_{n,l}^{(m)} e^{il \chi
 +i\frac{l}{|l|}\chi} \sbr{\frac{n! \cbr{\frac{ \kappa_0}{2}}^{|l|+1}}{\pi (n + |l|)!}}^{\frac{1}{2}} \sqrt{\tilde{\alpha}^{|l|}}\frac{\tilde{\alpha}\kappa_0r^{|l|+1}}{2} L_{n}^{|l|+1}e^{-\frac{\tilde{\alpha} \kappa_0 r^2}{2}}\\
 &+2\sum_{n=0}^{\infty}\lambda\tilde{\lambda}^{m}\tilde{\alpha}^{n+1}C_{n,l}^{(2n)}\cbr{\frac{e^{i\chi}}{2}+\frac{e^{-i\chi}}{2}} \sqrt{\frac{\kappa_0}{2\pi }} \frac{\tilde{\alpha}\kappa_0r}{2} L_{n}^{1}e^{-\frac{\tilde{\alpha} \kappa_0 r^2}{2}}\,.
\end{aligned}
\end{equation} 
Now we make a change on the dummy indices of the sums. In particular, for the first term we define
$$
l'=l-\frac{l}{|l|}, \quad n'=n+1
$$
so that $|l'|=|l|-1$ and the condition $2n+|l|=m$ becomes
$$
2n'+|l'|=m+1\,.
$$
For the second term, we introduce instead
$$
l''=l+\frac{l}{|l|}, \quad n''=n
$$
so that $|l''|=|l|+1$ and the condition $2n+|l|=m$ becomes
$$
2n''+|l''|=m+1\,.
$$
We find
\begin{equation}
\label{eq:sincos2}
\begin{aligned}
 &-\lambda \cos \chi \partial_r P+\lambda \frac{\sin \chi}{r} \partial_{\chi} P=\\
 =&-\sum_{n'=1}^{\infty}\sum_{|l'|>0}\delta_{m+1,2n'+|l'|}\lambda\tilde{\lambda}^{m}\sqrt{\tilde{\alpha}^{m+|l'|+3}}C_{n'-1,|l'|+1}^{(m)} e^{il' \chi} \sbr{\frac{n'! \cbr{\frac{ \kappa_0}{2}}^{|l'|+1}}{\pi (n + |l'|)!}}^{\frac{1}{2}} \sqrt{\frac{n'\kappa_0}{2}}r^{|l'|}L_{n'}^{|l'|}e^{-\frac{\tilde{\alpha} \kappa_0 r^2}{2}} \\
 &- \sum_{n=0}^{\infty}\lambda\tilde{\lambda}^{2n+1}\tilde{\alpha}^{n+2}C_{n,1}^{(2n+1)}\frac{\kappa_0^{2}\sqrt{n+1}}{2\sqrt{\pi}}L_{n+1}^0e^{-\frac{\tilde{\alpha} \kappa_0 r^2}{2}}\\
 &+\sum_{n''=0}^{\infty}\sum_{|l''|>0}\delta_{m+1,2n''+|l''|}\lambda\tilde{\lambda}^{m}\sqrt{\tilde{\alpha}^{m+|l''|+3}}C_{n'',|l''|-1}^{(m)} e^{il'' \chi} \sbr{\frac{n''! \cbr{\frac{ \kappa_0}{2}}^{|l''|+1}}{\pi (n'' + |l''|)!}}^{\frac{1}{2}} \sqrt{\frac{(n''+|l''|)\kappa_0}{2}}r^{|l''|} L_{n''}^{|l''|}e^{-\frac{\tilde{\alpha} \kappa_0 r^2}{2}}\,.
\end{aligned}
\end{equation} 
The second term on the r.h.s. of the above equation comes from the $l=\pm1$ contributions to the first term on the r.h.s. of Eq.~\eqref{eq:sincos1}. Both the second and the third term on the r.h.s. of that equation contribute to the third term on the r.h.s. of Eq.~\eqref{eq:sincos2}.
We can drop all the prime symbols of the dummy indices and get,  recalling Eqs.~\eqref{eq:normalcn02n} and~\eqref{eq:iterc},
\begin{equation}
 \label{eq:sincos3}
\begin{aligned}
 &-\lambda \cos \chi \partial_r P+\lambda \frac{\sin \chi}{r} \partial_{\chi} P=\\
 =&\sum_{n=0}^{\infty}\lambda\tilde{\lambda}^{2n+1}\tilde{\alpha}^{n+2}C_{n+1,0}^{(2n+2)}\sqrt{2\kappa_0(n+1)}\frac{\kappa_0^{2}\sqrt{n+1}}{2\sqrt{\pi}}L_{n+1}^0e^{-\frac{\tilde{\alpha} \kappa_0 r^2}{2}}\\
 &+\sum_{n=0}^{\infty}\sum_{|l|>0}\delta_{m+1,2n+|l|}\lambda\tilde{\lambda}^{m}\sqrt{\tilde{\alpha}^{m+3}} \cbr{C_{n,|l|-1}^{(m)} \sqrt{\frac{(n+|l|)\kappa_0}{2}} - C_{n-1,|l|+1}^{(m)} \sqrt{\frac{n\kappa_0}{2}}  } \phi_{n,l} \\
 =&\sum_{n'''=0}^{\infty}\lambda\tilde{\lambda}^{n'''-1}\tilde{\alpha}^{n'''+1}C_{n''',0}^{(2n''')}\sqrt{2\kappa_0n'''}\frac{\kappa_0^{2}\sqrt{n'''}}{2\sqrt{\pi}}L_{n'''}^0e^{-\frac{\tilde{\alpha} \kappa_0 r^2}{2}}\\
 &+\sum_{n=0}^{\infty}\sum_{|l|>0}\delta_{m''',2n+|l|}\lambda\tilde{\lambda}^{m'''-1}\sqrt{\tilde{\alpha}^{m'''+2}} C_{n,l}^{(m''')}[\kappa_0m'''+l^2]\phi_{n,l} \\
\end{aligned}
\end{equation} 
where two last changes of indices, $n+1 \to n'''$ for the first term in the r.h.s. and $m+1 \to m'''$ for the second one, were made. By dropping the prime symbols and merging the two expressions we finally get
\begin{equation}
 \label{eq:sincos4}
 -\lambda \cos \chi \partial_r P+\lambda \frac{\sin \chi}{r} \partial_{\chi} P=\sum_{m=0}^{\infty}\lambda\tilde{\lambda}^{m-1}\sqrt{\tilde{\alpha}^{m+2}}\sum_{2n+|l|=m}C_{n,l}^{(m)}\sbr{m\kappa_0 +l^2}\phi\,.
\end{equation}

Putting together Eqs.~\eqref{eq:termt},~\eqref{eq:termr},~\eqref{eq:termchi2} and~\eqref{eq:sincos4}, one finally obtains 
\begin{equation}
 \label{eq:fpansatz}
 \begin{aligned}
 &\sum_{m=0}^{\infty}\tilde{\lambda}^{m}\sqrt{\tilde{\alpha}^{m+2}}\sum_{2n+|l|=m}C_{n,l}^{(m)}( \kappa_0)\phi_{n,l}(\sqrt{\tilde{\alpha}}r,\chi|\kappa_0)\sbr{\frac{\dot{\tilde{\alpha}}}{2\tilde{\alpha}}-\kappa+\tilde{\alpha} \kappa_0}\cbr{|l|+2-\tilde{\alpha} \kappa_0 r^2-\tilde{\alpha} \kappa_0 r^2\frac{L_{n-1}^{|l|+1}}{L_{n}^{|l|}}}+\\
 &+\sum_{m=0}^{\infty}\tilde{\lambda}^{m}\sqrt{\tilde{\alpha}^{m+2}}\sum_{2n+|l|=m}C_{n,l}^{(m)}(\kappa_0)\phi_{n,l}(\sqrt{\tilde{\alpha}}r,\chi|\kappa_0)\sbr{\frac{\dot{\tilde{\alpha}}}{2\tilde{\alpha}} + \frac{\dot{\tilde{\lambda}}}{\tilde{\lambda}}+\tilde{\alpha} \kappa_0 -\kappa_0\frac{\lambda}{\tilde{\lambda}}}m+\\
 &+\sum_{m=0}^{\infty}\tilde{\lambda}^{m}\sqrt{\tilde{\alpha}^{m+2}}\sum_{2n+|l|=m}C_{n,l}^{(m)}(\kappa_0)\phi_{n,l}(\sqrt{\tilde{\alpha}}r,\chi|\kappa_0)\sbr{\frac{\lambda}{\tilde{\lambda}}-1}l^2=0\,.
 \end{aligned}
\end{equation}
For the above equation to hold for every choice of $r$ and $\chi$, the expressions in square parenthesis needs to vanish. One has therefore
\begin{subequations}
\label{eq:finalresult}
 \begin{eqnarray}
 \kappa &= &\frac{\dot{\tilde{\alpha}}}{2\tilde{\alpha}}+ \kappa_0\tilde{\alpha}\\
\dot{\lambda}&=&-\sbr{\kappa-\kappa_0}\lambda\,\\
\tilde{\lambda}&=&\lambda\,.
\end{eqnarray} 
\end{subequations}
Integrating the second equation and recalling condition~\eqref{eq:condition}, one recovers
Eqs.~(\eqmain) of the main text.

\section{Average work and $\av{r^2}$}
\label{sec:appw}
In this section, we compute the average confining work performed by the external forces during the protocols described in the main text. To this end, we will also compute the explicit expression of $\av{r^2}$.
The average work is defined as
\begin{equation}
    \av{W}=\int_0^{t_f}dt\,\int_0^{2\pi}d\chi\,\int_0^{\infty}dr\,r\, \partial_tU(r|\kappa)P(r,\chi|\tilde{\alpha}, \lambda,\kappa_0)=\frac{D_t}{2\mu}\int_0^{t_f}dt\,\dot{\kappa}\,\av{r^2}\,,
\end{equation}
where we have expressed the external potential in terms of the rescaled variables:
\begin{equation}
   U=\frac{k \rho^2}{2}=\frac{D_t \kappa r^2}{2\mu} \,.
\end{equation}
Let us first compute $\av{r^2}$.
By taking into account Eq.~\eqref{eq:pdfansatz} and recalling that
\begin{equation}
    \int_0^{2\pi}d \chi e^{il \chi}= 2\pi \delta_{l,0}
\end{equation}
we get
\begin{equation}
    \av{r^2}=2\pi \sum_{n=0}^{\infty}\lambda^{2n} \tilde{\alpha}^{n+1}C_{n,0}^{(2n)}\sqrt{\frac{ \kappa_0}{2 \pi}}\int_0^{\infty}dr\,r^3 e^{-\frac{\tilde{\alpha} \kappa_0 r^2}{2}}L_n^0\cbr{\frac{\tilde{\alpha} \kappa_0 r^2}{2}}\,.
\end{equation}

By making use of Rodrigues formula Eq.~\eqref{eq:rodrigues}, we can rewrite the integral as
\begin{equation}
    \begin{aligned}
    \int_0^{\infty}dr\,r^3 e^{-\frac{\tilde{\alpha} \kappa_0 r^2}{2}}L_n^0\cbr{\frac{\tilde{\alpha} \kappa_0 r^2}{2}}=&\frac{2}{\tilde{\alpha}^2 \kappa_0^2}\int_0^{\infty}dx\,x e^{-x}xL_n^0(x)\\
    =&\frac{2}{\tilde{\alpha}^2 \kappa_0^2}\int_0^{\infty}dx\,\frac{x}{n!} \frac{d^n}{dx^n}\cbr{e^{-x}x^n}\\
    =&\frac{2}{\tilde{\alpha}^2 \kappa_0^2}\cbr{\delta_{n,0}-\delta_{n,1}}\,,
    \end{aligned}
\end{equation}
leading to
\begin{equation}
\begin{aligned}
\label{eq:r2}
     \av{r^2}&=2\pi\frac{2\sqrt{2 \pi}}{\tilde{\alpha} \kappa_0\sqrt{\kappa_0}}\cbr{C_{0,0}^{(0)}-\lambda^2 \tilde{\alpha} C_{1,0}^{(2)}}\\
    &=\frac{1}{\kappa_0} \cbr{\frac{2}{\tilde{\alpha}}+\frac{ \lambda^2}{\kappa_0+1}}
\end{aligned}
\end{equation}
and therefore
\begin{equation}
     \av{W}=\frac{D_t}{2\mu \kappa_0}\int_0^{t_f}dt\, \dot{\kappa}\cbr{\frac{2}{\tilde{\alpha}}+\frac{ \lambda^2}{\kappa_0+1}}\,.
\end{equation}

\section{Characterization of the minimal time protocol}
\label{sec:sm_minimaltime}
\subsection{Minimal time dependence on system's parameters}
\begin{figure}
    \centering
    \includegraphics[width=0.99\linewidth]{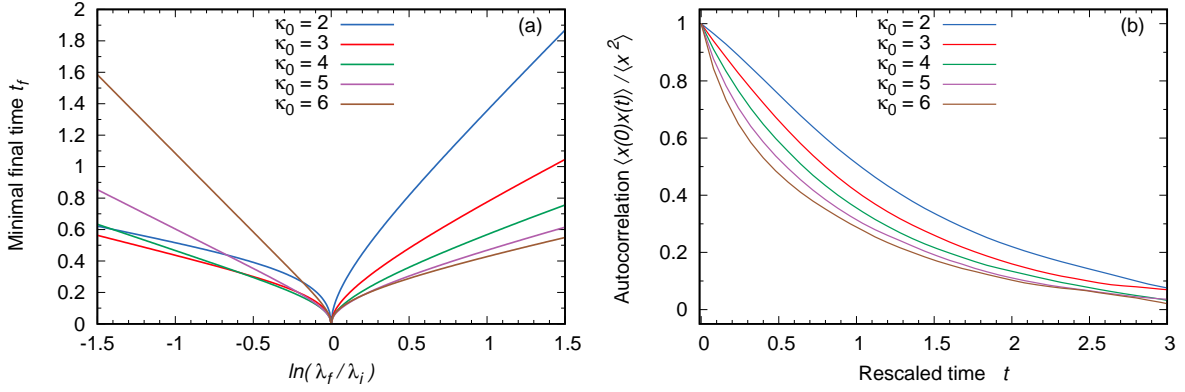}
    \caption{Panel (a): Minimal time $t_f^{min}$ for the protocol studied in the main text as a function of the ratio $\lambda_f/\lambda_i$. Both increasing and decreasing activities are taken into account. Different colors correspond to different values of $\kappa_0$. The constraint $1\le \kappa(t) \le 7 $ is assumed. Panel (b): Autocorrelation function $\langle x(0)x(t) \rangle/\langle x^2 \rangle$ in the stationary state, $x(t)$ being the projection on the $x$ axis of the position of the particle at time $t$, obtained as an average over $10^4$ independent simulations of the process, with $\lambda=5$; see Section \ref{sec:simul} of this SM for details on the simulations. For the considered values of $\kappa$, typical relaxation times are of order $1$. }
    \label{fig:tf}
\end{figure}

As discussed in the main text, the minimal time for the   class of protocols studied can be computed by solving numerically Eqs.~(\eqtfa) and (\eqtfb). In Fig.~\ref{fig:tf}(a), we report the dependence of the solution on the boundary conditions for the control parameters. Fig.~\ref{fig:tf}(b) shows, for comparison, the position autocorrelation functions measured  from numerical simulations, from which the typical relaxation times of the process can be inferred.
The curves in Fig.~\ref{fig:tf}(a) are determined by a non-trivial combination of two effects. On the one hand, larger values of $\kappa_0$ correspond, in general, to faster dynamics [see Fig.~\ref{fig:tf}(b)]. On the other hand, the variation rate of $\lambda$ is proportional to $\kappa_0-\kappa$; if $\kappa_0$ is very close to the upper bound $\kappa_{+}$, the decrease of $\lambda$ will be slow. The former effect is dominant when the desired activity variation
is small. The latter becomes more relevant in the limit of large $|\ln(\lambda_f/\lambda_i)|$.

\subsection{Approaching the optimal protocol by polynomials}

\begin{figure}
    \centering
    \includegraphics[width=0.59\linewidth]{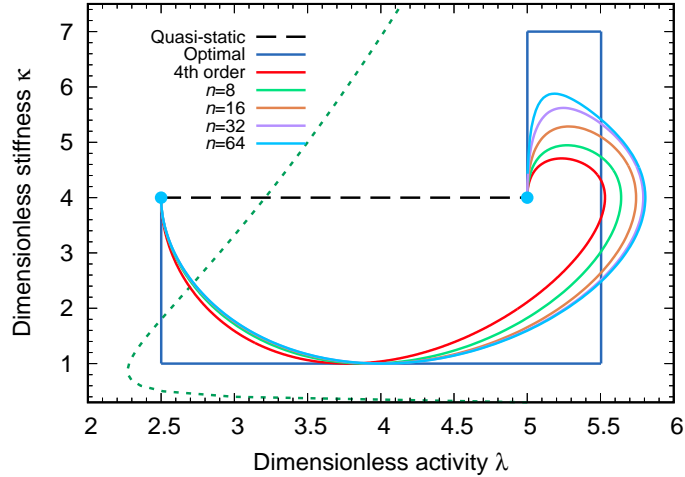}
    \caption{Examples of protocols that can be obtained by choosing the functional form Eq.~\eqref{eq:ffalphapoly}, with different values of the exponent $n$, and imposing the minimization of the total time constrained to the $1<\kappa<7$ condition. By just adding a single degree of freedom to the polynomial functional form discussed in the main text, it is already possible to notice that the protocol approaches the optimal one. The dotted curve separates the ``active-like'' from the ``passive-like'' region, as defined in the main text.}
    \label{fig:oplimit1}
\end{figure}

The proposed protocol is the fastest process that we can devise, among the ones that are compatible with our ansatz for the solution of the Fokker-Planck equation and with the chosen constraints for the stiffness. This consequently means that by considering polynomial forms for $\tilde{\alpha}(t)$ of higher and higher order, subjected to the same constraints, one should approach the optimal protocol upon further minimizing $t_f$.
Yet, it is not straightforward to see this explicitly, since the constrained optimization is already quite involved (even numerically) when just a couple of free parameters are at play. Still, quite a clear hint of this limiting behaviour can be obtained by considering the functional form
\begin{equation}
\label{eq:ffalphapoly}
    \tilde{\alpha}(t)=F_n(t)=1+\tilde{\alpha}_2 t^2+\tilde{\alpha}_3 t^3+\tilde{\alpha}_4 t^4+\tilde{\alpha}_n t^n\,.
\end{equation}
In the absence of the last term, we have the fourth-order polynomial discussed in the main text. $t_f$ can be chosen as the minimal one that guarantees $\kappa_{-} < \kappa < \kappa_{+}$ (red curve in Fig.~\ref{fig:oplimit1}).
If we also include the larger order term $\tilde{\alpha}_n t^n$ with $n >4$, then we get protocols that look closer to the optimal one, once the minimization condition are enforced to determine the optimal values of the coefficients. As shown in  Fig.~\ref{fig:oplimit1}, when $n$ increases, the path in the parameter space approaches the qualitative behavior of the optimal protocol, also in the $\kappa>\kappa_0$ phase. This improvement is obtained by just adding one degree of freedom to the considered functional form of $\tilde{\alpha}(t)$. Closer to optimal results are expected to be obtained if more parameters are considered, at the price of a possibly demanding numerical optimization of the coefficients.

\subsection{Approximating the optimal protocol by continuous-in-time functions}

\begin{figure}
    \centering
    \includegraphics[width=0.99\linewidth]{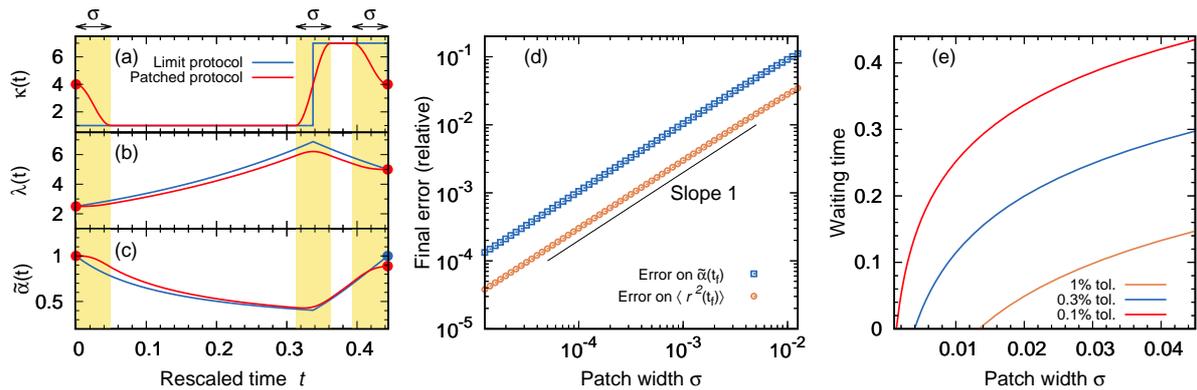}
    \caption{Continuous ``patched'' protocol mimicking the optimal one. In panels (a-c), the relevant parameters of the dynamics are reported for the optimal (blue curves) and the patched (red curves) protocol. The yellow-shaded areas correspond to the time intervals when the patched protocol $\kappa_{patch}(t)$ is different from $\kappa(t)$. Here $\sigma=0.05$.} Panel (d) shows  how the relative errors on $\tilde{\alpha}(t_f)$ and $\av{r^2(t_f)}$ in the final state depend on the patch width $\sigma$. The black straight line is a guide for the eyes, suggesting a linear dependence of the error with $\sigma$. Finally panel (e) reports the waiting time needed \textit{after} the conclusion of the approximated protocol to achieve the target state, as a function of $\sigma$. As discussed in the text, the target state is considered reached when the value of $\av{r^2}$ matches the target one, within the given level of tolerance.
    \label{fig:oplimit2}
\end{figure}

At a practical level, there are simpler ways to approximate the limit optimal protocol by continuous functions, as required by real experimental setups. The optimal protocol
\begin{equation}
    \kappa(t)=\kappa_{-}\Theta\sbr{(t^{\star}-t)t}+\kappa_{+}\Theta\sbr{(t-t^{\star})(t_f-t)},
\end{equation}
where $\Theta$ denotes the Heaviside step function,
can be made continuous by introducing suitable ``patches'' of the form
\begin{equation}
\kappa_{patch}(t)=A + B \cos\cbr{\frac{\pi t}{\sigma}+C}
\end{equation}
in place of the discontinuities. Here $\sigma$ is a parameter accounting for the time one needs for the change, see Fig.~\ref{fig:oplimit2}(a), while $A$, $B$ and $C$ need to be adjusted in order for $\kappa_{patch}$ to match the value of the optimal protocol before and after the considered discontinuity point. In the limit $\sigma \to 0$ the optimal protocol is recovered. Once a functional form for $\kappa(t)$ is known, $\tilde{\alpha}(t)$ and $\lambda(t)$ can be derived by exploiting Eqs.~(\eqmain) of the main text.
Figure~\ref{fig:oplimit2}(a-c) shows how the process changes as a consequence of this deviation from the optimal prescription. The net result is that, at the end of the process, the value of $\tilde{\alpha}$ will not be equal to $1$, as it should be in the final state, meaning that the desired stationary state has not been reached yet. The error $|1-\tilde{\alpha}(t_f)|$ on the final state can be made arbitrary small, provided that patched protocols with sufficiently small $\sigma$ are realized. In Fig.~\ref{fig:oplimit2}(d) this ``distance'' from the final stationary state is shown as a function of $\sigma$. In the same panel also the relative error
$$
\Bigg|\frac{\av{r^2(t_f)}-\av{r^2}_s}{\av{r^2}_s}\Bigg|
$$
on the variance of the radial position is shown, where $\av{r^2}_s$ denotes its value in the target stationary state.

We can ask ourselves how long it will take to the system to reach the actual target state, once the patched protocol has been performed. This problem can be addressed by noticing that at the end of the  protocol the pdf of the system is still of the form~\eqref{eq:statpdf}; indeed, by construction the proposed dynamics preserves that functional shape for the pdf. As a consequence, we can repeat the considerations in Section III and see that Eqs.~\eqref{eq:finalresult} are still valid. In particular one gets
\begin{equation}
    \dot{\tilde{\alpha}}=2\kappa_0(1-\tilde{\alpha})\tilde{\alpha}\,,
\end{equation}
which leads to the solution
\begin{equation}
\label{eq:alphapost}
    \tilde{\alpha}(t)=\frac{e^{2\kappa_0(t-t_f)}\tilde{\alpha}(t_f)}{e^{2\kappa_0(t-t_f)}\tilde{\alpha}(t_f)-\tilde{\alpha}(t_f)+1}, \quad t\ge t_f\,.
\end{equation}
The actual stationary state $\tilde{\alpha}(t)=1$ is reached asymptotically in time, for $t \to \infty$. However, in experimental situations where the above continuous approximation is relevant, a finite tolerance on the results is accepted (given, e.g., by the finite resolution of the measuring devices). It is thus interesting to compute the time needed in order to reach a given accuracy on a typical observable of the system, e.g. $\av{r^2}$.

Equation~\eqref{eq:alphapost} can be plugged in~\eqref{eq:r2} to get the evolution of $\av{r^2}$. In this way, we can estimate how long one needs to wait, after the end of the protocol, until $\av{r^2(t)}$ is equal to the target value, within a given tolerance. The results of this analysis are shown in Fig.~\ref{fig:oplimit2}(e), where the waiting time is plotted as a function of $\sigma$, for given values of the tolerance on $\av{r^2}$. Let us imagine, for instance, that the experimental setup in use can approximate discontinuous jumps via patches with $\sigma=0.01$. Then in order to reach the final state within a tolerance of $0.1\%$ on the value of $\av{r^2}$, an additional time of $\simeq0.25$, in rescaled units, will be needed at the end of the protocol; on the other hand, if we settle for a $1\%$ accuracy, no extra time will be needed, as the final state provided by the protocol already satisfies the requested tolerance. For a given value of the accuracy, the waiting time is a monotonously increasing function of $\sigma$: put in a different way, the experimental ability of realizing processes that are closer to discontinuous jumps is rewarded by a lower (or even negligible) waiting time after the end of the protocol. 

\section{Details on the numerical simulations}
\label{sec:simul}
The numerical simulations for this work have been obtained by integrating the Langevin equation for the process with an Euler-Maruyama scheme~\cite{kloeden1992stochastic}.
 The state of the system is described at each time by the vector $(x=r\cos\varphi,y=r\sin\varphi,\theta)$, which is updated according to the following Euler-Maruyama discretization of the Langevin equation Eq.~(1) in rescaled units:
\begin{equation}
\begin{aligned}
    x(t+\Delta t)&=x(t)+\cbr{\lambda(t)\cos \theta(t) + \kappa(t)x(t)}\Delta t + \sqrt{2 \Delta t}w_x(t)\\
    y(t+\Delta t)&=y(t)+ \cbr{\lambda(t)\sin \theta(t) + \kappa(t)y(t)}\Delta t+\sqrt{2 \Delta t}w_y(t)\\
    \theta(t+\Delta t)&=\theta(t)+\sqrt{2 \Delta t}w_{\theta}(t)
\end{aligned}
\end{equation}
The integration time-step $\Delta t$ is $10^{-4}$.

The simulation proceeds as follows. First, $N$ independent positions are extracted according to a Gaussian distribution with unit variance; the $N$ particles are evolved with fixed values of $\kappa=\kappa_0$ and $\lambda=\lambda_i$ for a time $t=10$ (the typical autocorrelation time of the process being $\simeq 1$, see Fig.~\ref{fig:tf}). This is the initial time at which the protocol starts. Then $\kappa(t)$ and $\lambda(t)$ are evolved following the prescribed protocol, and the values of the position are recorded.
At each time step it is possible then to compute characteristic observables of the dynamics, like the variance of the radial position $\av{r^2}$:
\begin{equation}
\av{r^2(t)}=\frac{x(t)^2+y(t)^2}{N}\,,
\end{equation}
where $N$ is the number of independent realizations of the process.


\bibliography{biblio}